\documentclass[aip,reprint]{revtex4-1}

\draft 
\usepackage[pdftex]{graphicx}

\begin{document}

\title{Decoupling of epitaxy related trapping effects in AlGaN/GaN metal-insulator semiconductor high electron mobility transistors}


\author{M.~Huber}
\affiliation{Infineon Tachnologies Austria AG, Siemensstrasse 2, A9500 Villach, Austria}
\affiliation{Institut f\"ur Halbleiter-und-Festk\"orperphysik, Johannes Kepler University, Altenbergerstr. 69, A-4040 Linz, Austria}

\author{G.~Pozzovivo}
\affiliation{Infineon Tachnologies Austria AG, Siemensstrasse 2, A9500 Villach, Austria}

\author{I.~Daumiller}
\affiliation{Infineon Tachnologies Austria AG, Siemensstrasse 2, A9500 Villach, Austria}

\author{G.~Curatola}
\affiliation{Infineon Tachnologies Austria AG, Siemensstrasse 2, A9500 Villach, Austria}

\author{L.~Knuuttila}
\affiliation{Infineon Tachnologies Austria AG, Siemensstrasse 2, A9500 Villach, Austria}

\author{M.~Silvestri}
\affiliation{Infineon Tachnologies Austria AG, Siemensstrasse 2, A9500 Villach, Austria}

\author{A. Bonanni}
\affiliation{Institut f\"ur Halbleiter-und-Festk\"orperphysik, Johannes Kepler University, Altenbergerstr. 69, A-4040 Linz, Austria}

\author{A.~Lundskog}
\affiliation{Infineon Tachnologies Austria AG, Siemensstrasse 2, A9500 Villach, Austria}






\keywords{MOVPE, Trapping, MIS-HEMT}

\begin{abstract}
%
%
%
The decoupling of epitaxial factors influencing on the dynamic instabilities of AlGaN/GaN metal-insulator semiconductor high electron mobility transistors is investigated. Three different sets of samples have been analyzed by means of dynamic instabilities in the threshold voltage (V$_{\mathrm{th}}$ shift). Secondary ion mass spectroscopy, steady-state photoluminescence (PL) measurements have been performed in conjunction with electrical characterization. The device dynamic performance is found to be significantly dependent on both the C concentration close to the channel as well as on the distance between the channel and the higher doped C region. Additionally, we note that experiments studying trapping should avoid large variations in the sheet carrier density (N$_{\mathrm{s}}$). This change in the N$_{\mathrm{s}}$ itself has a significant impact on the V$_{\mathrm{th}}$ shift. This experimental trends are also supported by a basic model and device simulation. Finally, the relationship between the yellow luminescence (YL) and the band edge (BE) ratio and the V$_{\mathrm{th}}$ shift is investigated. As long as the basic layer structure is not changed, the YL/BE ratio obtained from steady-state PL is demonstrated to be a valid method in predicting trap concentrations in the GaN channel layer.
\end{abstract}

%
%

\maketitle   

\section{Introduction}
\begin{table*}[t]
  \caption{Overview of the studied sample sets, including information on the distance between the channel and the highly C doped multilayer structure region, C concentration in the UID-GaN layer and in the barrier overlayer as well as on the type of barrier used.}
\begin{tabular}{p{1cm}p{4cm}p{2cm}p{2cm}p{2cm}p{3.5cm}}
\hline\hline
Sample&Sample-set&UID thickness & Carbon UID & Carbon AlGaN barrier & AlGaN barrier type \\
&& [nm] & [$cm^{-3}$]& [$cm^{-3}$]&\\
\hline\hline
A&UID thickness series&1000&low $10^{-16}$ &low $10^{-17}$ &standard\\
B&UID thickness series&300&low $10^{-16}$ &low $10^{-17}$ &standard\\
C&UID thickness series&500&low $10^{-16}$ &low $10^{-17}$ &standard\\
D&UID thickness series&750&low $10^{-16}$ &low $10^{-17}$ &standard\\
\hline
E&UID impurity series&500&low $10^{-16}$ &low $10^{-17}$ &standard\\
F&UID impurity series&500&low $10^{-16}$ &low $10^{-17}$ &standard\\
G&UID impurity series&500&high $10^{-15}$ &low $10^{-17}$ &standard\\
H&UID impurity series&500&med $10^{-16}$ &low $10^{-17}$ &standard\\
I&UID impurity series&500&high $10^{-16}$ &low $10^{-17}$ &standard\\
J&UID impurity series&500&low $10^{-17}$ &low $10^{-17}$ &standard\\
\hline
K&Barrier series&500&low $10^{-16}$ &low $10^{-16}$ &thick / high Al\\
L&Barrier series&500&low $10^{-16}$ &low $10^{-16}$ &thick / medium Al\\
M&Barrier series&500&low $10^{-16}$ &low $10^{-16}$ &thick / low Al\\
N&Barrier series&500&low $10^{-16}$ &med $10^{-16}$ &thin / medium Al\\
O&Barrier series&500&low $10^{-16}$ &low $10^{-17}$ &very thick / very low Al\\
P&Barrier series&500&low $10^{-16}$ &low $10^{-16}$ &medium / medium Al\\
 \hline
 \hline
\end{tabular}
  \label{TBL1}
\end{table*}

The AlGaN/GaN material system is a fundamental building-block for the fabrication of power GaN on Si metal insulator semiconductor high electron mobility transistors (MIS-HEMTs) \cite{Kanamura2012,Ueda2014}. While Si is a widely used substrate for nitride-based HEMTs, complex multilayer structures are needed in order to keep a strained architecture and ensure low vertical leakage currents. One way to satisfy these requirements, is to introduce acceptor-like impurities, like e.g. Fe or C \cite{Silvestri2013,Poblenz2004}. While Fe is considered as non-suitable element for the Si front-end-of-line area, C could potentially fulfill this task \cite{Treidel2010}. 

For a functioning GaN based device, it is of high importance to optimize trapping related failures such as current collapse, dynamic on-resistance (R$_{\mathrm{DSON}}$) degradation and threshold voltage (V$_{\mathrm{th}}$) shifts. The location of the physical trapping sites and mechanisms causing dynamic instabilities in C compensated HEMT structures has been described in literature \cite{Bisi2015}. It was discussed, that these effects are typically increasing by the implementation of such low leakage current C compensated buffers \cite{Klein2001a,Uren2012,Wurfl2013,Verzellesi2014a}. On the other hand, Moens et. al. have recently shown that a high leakage current is not necessarily needed to achieve low dynamic instabilities \cite{Moens2015}. Others showed that a significant part of the trapping also takes place at the interface between the dielectric and the III-nitride semiconductor \cite{Lagger2012,Lagger2014}. Combinations of these mentioned mechanisms have also been reported \cite{Meneghesso2014}.

A typical signature of C doping of GaN is a broad yellow luminescence (YL) band around 2.2 eV. A clear attribution of the YL in GaN-related materials to a single defect has been matter of controversy for long \cite{Reshchikov2005}. Recently, the YL band was attributed to C substitution at the N site C$_{\mathrm{N}}$ and at the C$_{\mathrm{N}}$O$_{\mathrm{N}}$ complex \cite{Reshchikov2014}. Furthermore, the impact of the C$_{\mathrm{N}}$O$_{\mathrm{N}}$ complex located in the top epitaxial structure on the dynamic device instabilities was shown \cite{Huber2015}.

Although there are valid indications to minimize the C concentration to improve dynamic instabilities in AlGaN/GaN MIS-HEMTs, a conclusive picture of growth related epitaxial influence factors in the top epitaxial structure is still not available. In this work we investigate structural and impurity based influence factors on the dynamic instabilities caused by carrier trapping in the top epitaxial part. The major goal is a decoupling of the influence factors in the MIS-HEMT devices arising from epitaxial growth conditions close to the channel, as well as the impact from the channel properties itself.

\section{Experiment}
\begin{figure}[t]
\centering
\includegraphics[scale=0.48]{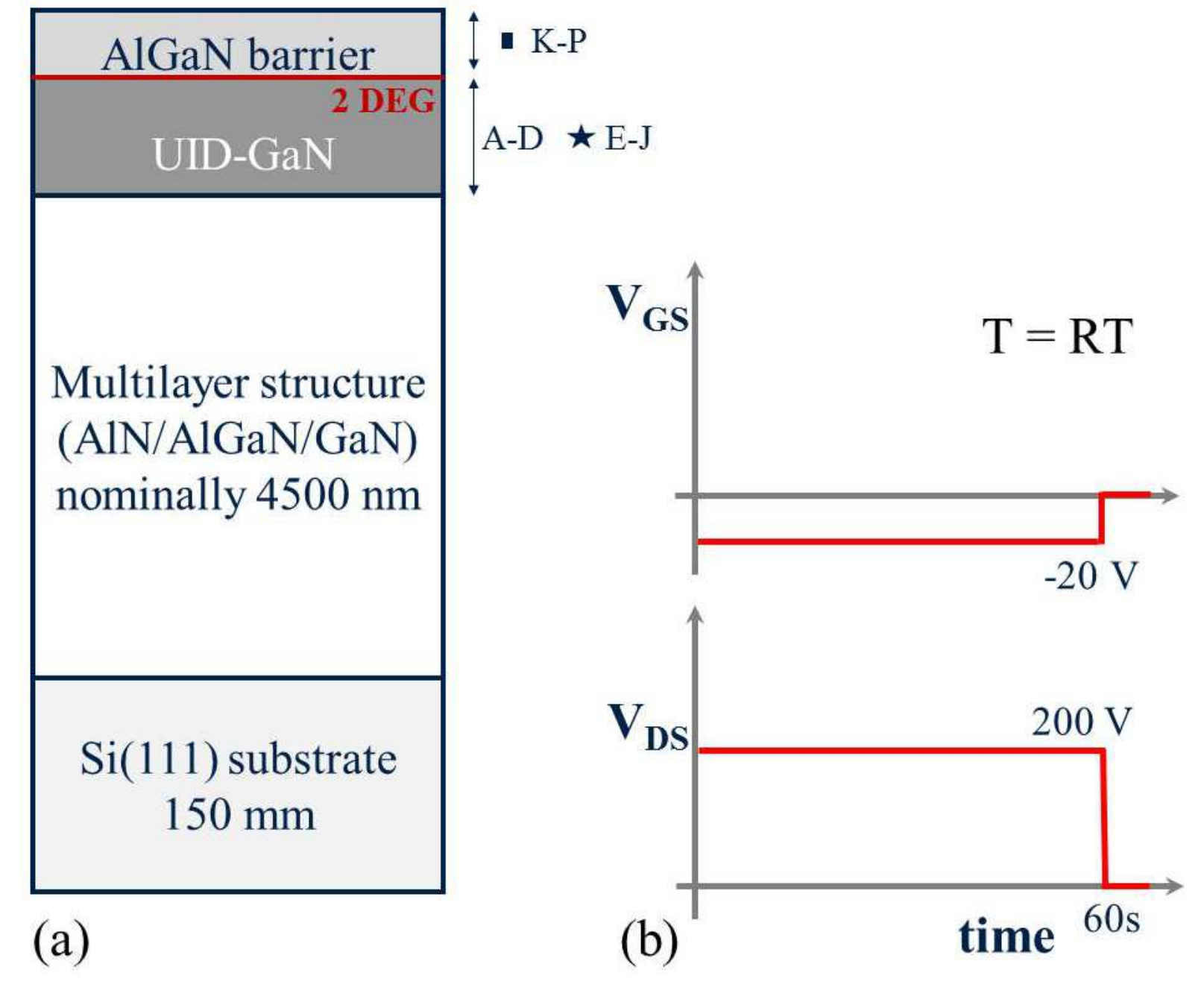}
\caption{(a) Sketch of the studied MIS-HEMT structure, where only the two upper most layers UID and barrier are varied over the samples series. Star: variation in C concentration, square: variation in Al content and arrows: variation in thickness.  (b) Schematic overview of the timing and voltage conditions for the V$_{\mathrm{th}}$ shift measurement.}
\label{FIG.1}
\end{figure}

In Table \ref{TBL1} an overview of the studied samples is shown. The samples can be divided in three sets  including respectively the unintenionally dopded (UID) impurity series (Samples A-D), the UID thickness series (Samples E-J) and the barrier series (Samples K-P). The nominal epitaxial stack is shown in Fig.\ref{FIG.1}(a). The MIS-HEMT devices consist of an AlN/AlGaN/GaN hetero-structure grown on Si(111) 150 mm substrates using an multi-wafer AIXTRON planetary metal organic vapor phase epitaxy (MOVPE) reactor. The heterostructure consists of a highly resistive multilayer AlN/AlGaN/GaN grown directly on the substrate, a subsequent unintenionally doped (UID)-GaN layer and an AlGaN barrier overlayer. The nominal thicknesses of the layers are 4500 nm, 500 nm and 20 nm, respectively. The AlGaN barrier overlayer and the UID-GaN layer form the two dimensional electron gas (2DEG) channel of the HEMT structure.

The impurity concentrations in the layers have been characterized using secondary ion mass spectroscopy (SIMS) analysis. Measurements have been performed in a Phi-Evans quadrupole secondary ion mass spectrometry system equipped with Cs primary ion source operated at a low range keV energy to determine the impurity concentrations in the UID-GaN layer.

The material characteristics are correlated with the electrical performances of the MIS-HEMT devices fabricated by using a complementary metal-oxide-semiconductor (CMOS) production line. The 2DEG sheet resistance of the standard AlGaN barrier is $\sim$460 Ohm/sq. Hall mobility and 2DEG density are 1800 cm$^2$/Vs and 7.5x10$^1$$^2$ cm$^-2$, respectively. The Au-free ohmic contacts consist of a Ti/Al based metal stack with R$_c$=0.5 Ohm$\cdot$mm. The gate length is 1 $\mu$m, while the gate-source distance and gate-drain distance are 1.5 $\mu$m and 12 $\mu$m, respectively. The rated breakdown voltage is 650 V at 10 nA/mm. The devices are passivated by using SiN and poly-imide. Electrical characterization includes various stress tests up to 200 V. For the threshold voltage shift, the devices have been subjected to a stepped OFF-state stress, keeping the gate source voltage (V$_{\mathrm{GS}}$ = -20 V) fixed. The stress pulse duration is set to one minute and the drain voltage to 200 V. The V$_{\mathrm{th}}$ is measured directly after applying the stress pulse (Fig.\ref{FIG.1}(b)).

\begin{figure}[t]
\centering
\includegraphics[scale=1.02]{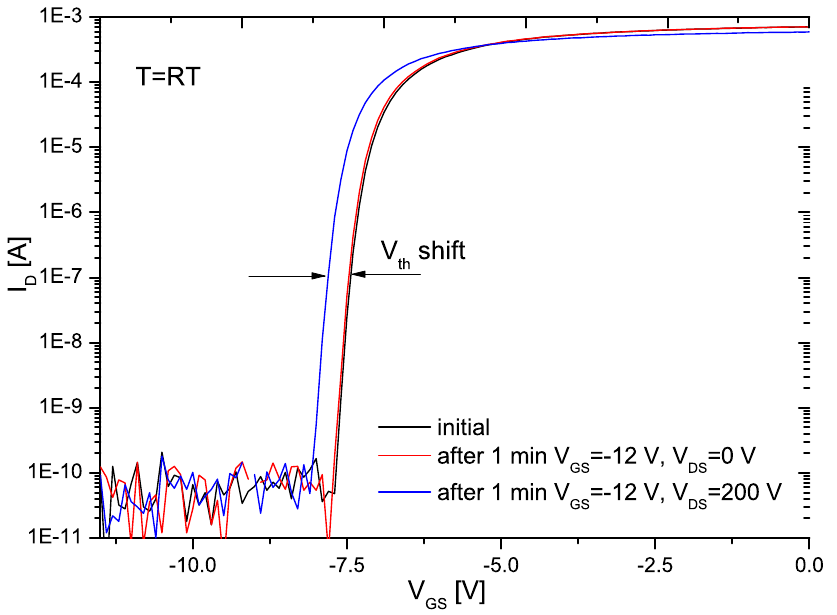}
\caption{Transfer curve plot indicating the V$_{\mathrm{th}}$ shift measurement for sample H at V$_{\mathrm{DS}}$=0 V (unstressed) and V$_{\mathrm{DS}}$=200 V (after stress).}
\label{FIG.2}
\end{figure}

The obtained structures have been characterized by room temperature photoluminescence (PL) measurements using a HeCd laser for excitation with a wavelength of 325 nm and a power density of 0.1 W/mm$^2$.

\section{Results and Discussion}
The results of the SIMS measurements for samples A-P are summarized in Table\ref{TBL1} and significant signals from C are observed \cite{Evans2007}. In the UID impurity series the variation of growth parameters is known to have a particularly large impact on the C concentrations.~\cite{Armstrong2006}. It is worth to recall, that via SIMS measurements only the overall concentration of C in the UID-GaN layer is obtained, while neither information on the electrically active C, nor the actual incorporation site of C in the lattice can be identified. Since the variation of the growth conditions in the UID thickness series has also an effect on the growth rate, the overall UID-GaN layer thickness has been kept constant by adjusting the deposition times. In the UID thickness series basically only the thickness of the UID GaN layer is varied between 300 nm and 1000 nm, with corresponding adjustment in the multilayer structure to ensure a constant total thickness. In the last sample set, the barrier series, the AlGaN overlayer composition and thickness have been changed. In every series the non varying part of the structures has been kept constant.

In order to study the dynamic instabilities of the devices, the threshold voltage shift with the application of a high-voltage OFF-state stress pulse have been studied. Due to trapping mechanisms, V$_{\mathrm{th}}$ is shifted to more negative values with respect to the initial measurement after the application of the stress pulses. This behavior was recently attributed to electron traps in the GaN channel that are emptied under measurement conditions and filled under stress conditions \cite{Meneghesso2014}.
The threshold voltage shift is given in percentage of the initial values (Eq.~\ref{eq1}):

\begin{equation}
\label{eq1}
V_{th} (shift) =\frac{V_{th} (stress)-V_{th} (initial)}{V_{th} (initial)}\cdot100\%\\
\end{equation}

A typical example of the shift in the threshold voltage from sample H is shown in the MIS-HEMT transfer curve reported in Fig.\ref{FIG.2}.

\begin{figure}[]
\centering
\includegraphics[scale=1]{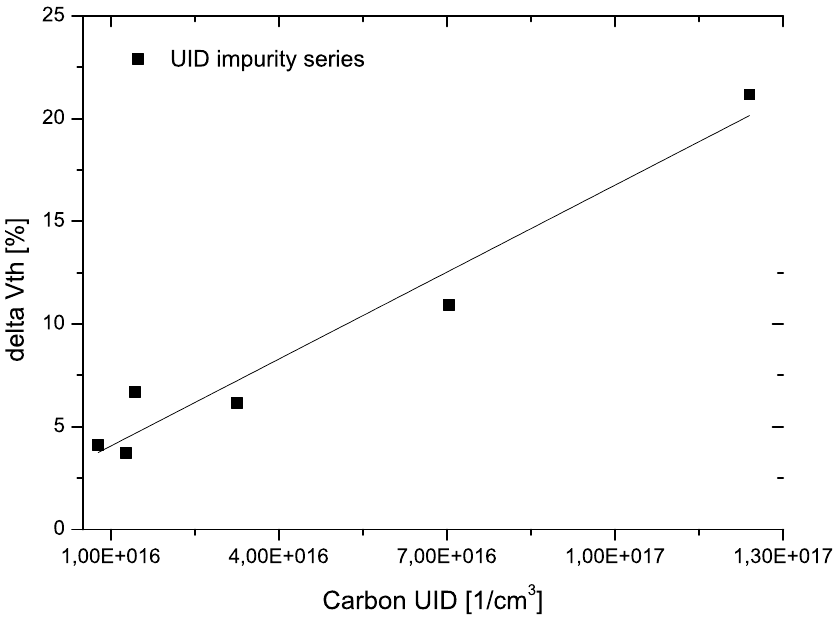}
\caption{V$_{\mathrm{th}}$ s shift for a device stressed at  V$_{\mathrm{DS}}$=200 V and V$_{\mathrm{GS}}$=-20 V plotted versus the carbon concentration in the UID GaN layer.}
\label{FIG.3}
\end{figure}

\subsection{Influence of the impurity concentration close to the channel}
For the impurity series (Sample A-D), the UID-GaN layer thickness has been fixed at 500nm and the C concentration in the UID-GaN has been varied by changing the MOVPE growth conditions. The resulting effect on the V$_{\mathrm{th}}$ s shift is shown in Fig.\ref{FIG.3} and a strong impact has been observed. No correlation to any other impurity observed from SIMS (hydrogen, silicon, oxygen) or positron annihilation spectroscopy (V$_{\mathrm{Ga}}$ s) has been found in this sample set, indicating a C related impurity complex in the UID-GaN layer to be the source of trapping \cite{Huber2015}. The observed results are consistent with the expectation, since the UID-GaN layer is directly beneath the 2DEG, electrons are easily trapped in any trap-states present. The C concentration in the UID-GaN layers of the other series are in the low $10^{-16}$ range. This is in fact in contrast with the findings in the UID impurity series and indicating still other contributions to V$_{\mathrm{th}}$ s shift in the MIS-HEMT structures.

\begin{figure}[]
\centering
\includegraphics[scale=0.95]{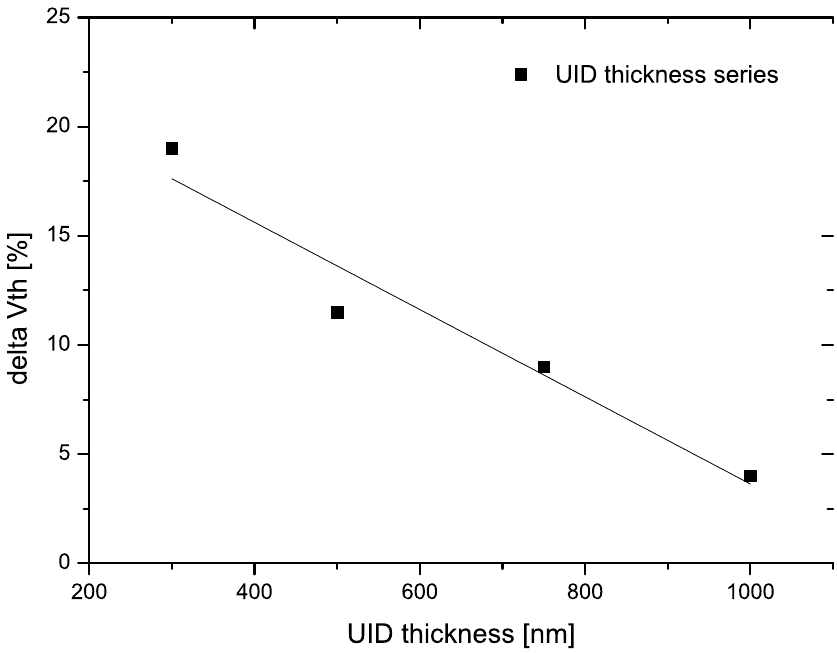}
\caption{V$_{\mathrm{th}}$ s shift for a device stressed at V$_{\mathrm{DS}}$=200 V and V$_{\mathrm{GS}}$=-20 V plotted over the distance from the channel to multilayer structure.}
\label{FIG.4}
\end{figure}

For the barrier series (samples K-P), the barrier layer thickness and Al content has been varied by changing the MOVPE growth conditions. The UID-GaN layer thickness has been fixed at 500 nm and the multilayer structure beneath has also not been changed. However, in the variations studied, a conclusive influence to none of the species measured with SIMS is observed. It must be recalled, that the AlGaN barrier layer is located at the surface and mere 20 nm thin, making quantitative SIMS measurements quite challenging.

\subsection{Influence of the distance between the channel and the higher doped C multilayer structure}
To answer the question, on whether the distance to the C doped multilayer structure also contributes to carrier trapping, the thickness of the UID-GaN layer has been varied in correlation with the multilayer structure thickness to maintain a fixed total thickness. This thickness determines the distance between the 2DEG and the highly C doped multilayer structure, which is in general expected to be a significant trapping site \cite{Wurfl2013,Verzellesi2014a,Moens2015}. In Fig.\ref{FIG.4} the V$_{\mathrm{th}}$ shift versus the UID-thickness is plotted. Up to an UID-GaN layer thickness of 1$\mu m$, the V$_{\mathrm{th}}$ shift has been found to continuously decrease with no sign of saturation. This indicates that the dynamic instabilties are a function of the distance between the channel and the C doped multilayer structure. As the distance of the UID-GaN layer is increased, the depletion during OFF-state stress in the multilayer structure decreases, causing fewer electrons to actually enter these trap-sites. It must be said, that this behavior could also be a function of the total thickness of the high doped multilayer structure and hence of the absoltue value of the leakage current in the structure. Further investigations are needed in order to draw a clear picture.

\subsection{Influence of the channel and 2DEG properties}
\begin{figure}[]
\centering
\includegraphics[scale=0.95]{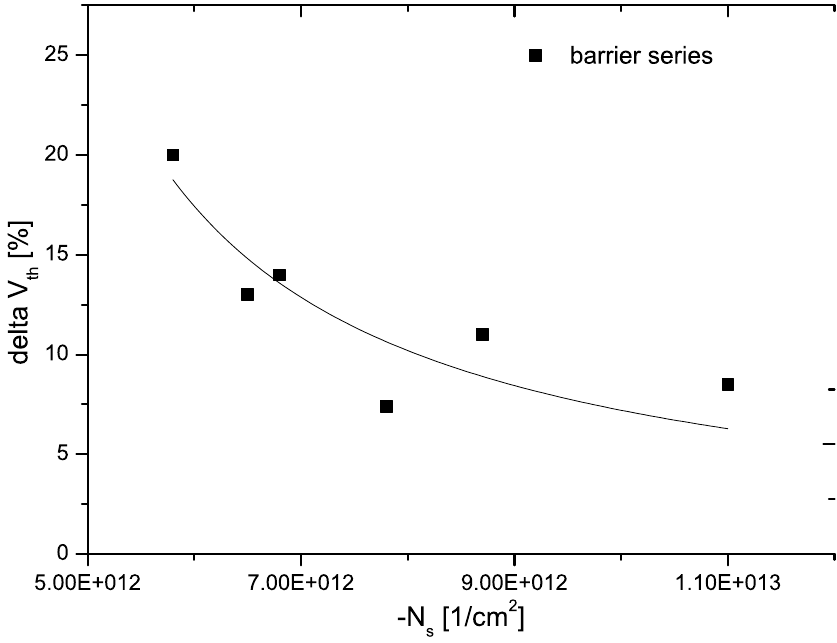}
\caption{V$_{\mathrm{th}}$ shift for a device stressed at V$_{\mathrm{DS}}$=200 V and V$_{\mathrm{GS}}$=-20 V plotted versus the sheet carrier density of the channel.}
\label{FIG.5}
\end{figure}

As expected, the above two sample-sets have virtually identical value of sheet carrier density (N$_{\mathrm{s}}$) in all the samples. In the third sample set, barrier thickness and aluminum content are varied while keeping all other parameters fixed. This gives a large variation in the N$_{\mathrm{s}}$ and the resulting effect on the V$_{\mathrm{th}}$ shift is shown in Fig.\ref{FIG.5} indicating that the V$_{\mathrm{th}}$ shift is lower the higher the N$_{\mathrm{s}}$. The V$_{\mathrm{th}}$ shift results from sample K of this series are not considered as relevant, as the threshhold voltage of this MIS-HEMT is very high (V$_{\mathrm{th}}$= -16 V), making the OFF-state stress with V$_{\mathrm{GS}}$=-20 V remarkably different from the other variations studied in this series.

\subsubsection{Free electron density with constant trap level}
\begin{figure}[]
\centering
\includegraphics[scale=0.97]{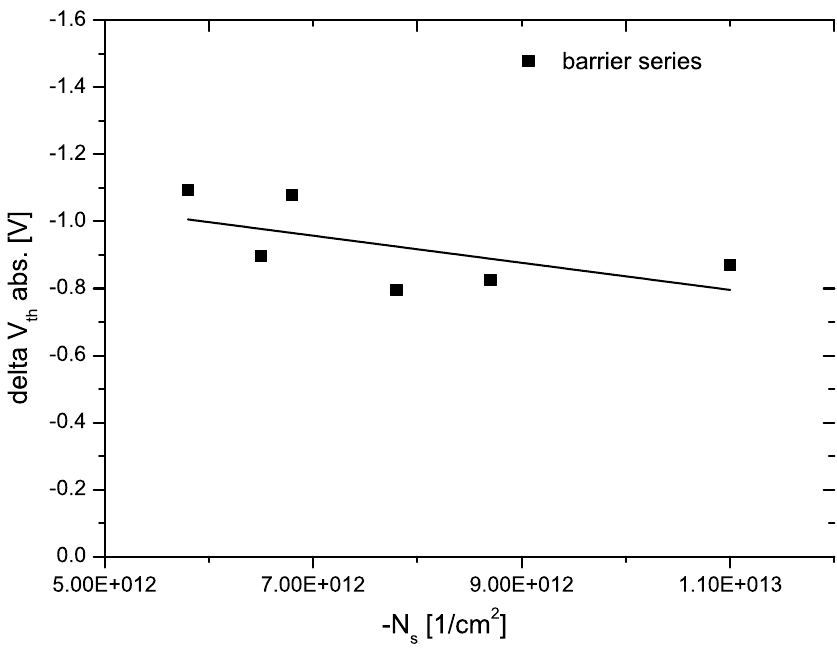}
\caption{Absolute value of the V$_{\mathrm{th}}$ shift in Volts for a device stressed at V$_{\mathrm{DS}}$=200 V plotted on top of the sheet carrier density of the channel.}
\label{FIG.6}
\end{figure}

The lower V$_{\mathrm{th}}$ shift values for higher N$_{\mathrm{s}}$ may be explained by the fact that a change in the free electron density is achieved while keeping the trap density unchanged. One possible option to test this hypothesis, is to consider the trap density itself as a function of the absolute value of the V$_{\mathrm{th}}$ shift. In Fig.\ref{FIG.6} this absolute value of the V$_{\mathrm{th}}$ shift in Volts is shown. As it can be seen the amount of charge trapped in the OFF-state conditions is virtually constant at $\sim$-0.9 V. As a result of the increased N$_{\mathrm{s}}$, trapping is considered to have a lesser effect on the V$_{\mathrm{th}}$ of the MIS-HEMT device. 

\subsubsection{AlGaN barrier thickness}
Additionally, it is generelly expected, that a thicker AlGaN barrier should reduce the effect of surface trapping and again changes the electric field during the OFF-state stress. However, as the thickness variation in the barrier series already causes a change in N$_{\mathrm{s}}$ and hence in the absolute value of the V$_{\mathrm{th}}$, this effect could not be studied without cross-influences. 

\subsubsection{1D calculation using linear model}
The findings relating the influence of the channel and 2DEG properties are not fully understood at the moment and will be a matter of further investiagtions. As a starting point of this discussion, a simple 1D formula is used to assess impact of the N$_{\mathrm{s}}$ value and the number of trap centers on the resulting device V$_{\mathrm{th}}$:

\begin{equation}
\label{eq2}
V_{th}\approx\frac{(N_{s}+N_{trap}) \cdot q \cdot (d_{SiN}+d_{Al})}{\epsilon_{0} \cdot \epsilon_{r}}\\
\end{equation}

with d$_{\mathrm{SiN}}$= 33 nm as the thickness of the gate nitride and d$_{\mathrm{Al}}$= 20 nm as the thickness of the AlGaN barrier, the relative permittivity ($\epsilon$$_{\mathrm{r}}$), the electric constant ($\epsilon$$_{\mathrm{0}}$) and the elementary charge (q). The amount of electrons trapped during OFF-state stress is approximated by summation of N$_{\mathrm{trap}}$. For the modelling of the UID impurity series: A varying number of trap centers is assumed to change from $10^{-11}$ cm$^{\mathrm{-2}}$ to $10^{-12}$ cm$^{\mathrm{-2}}$ similar to the relative variation of the C content in the UID impurity series (Table \ref{TBL1}). The slight variation in the N$_{\mathrm{s}}$ obtained from the experiment is also included in the model acting as further input parameter. For the modelling of the barrier series: A fixed number of trap centers is assumed at $8\cdot10^{-11}$ cm$^{\mathrm{-2}}$ together with input of the increasing value of N$_{\mathrm{s}}$ according to the experimental data. 

\begin{figure}[]
\centering
\includegraphics[scale=0.95]{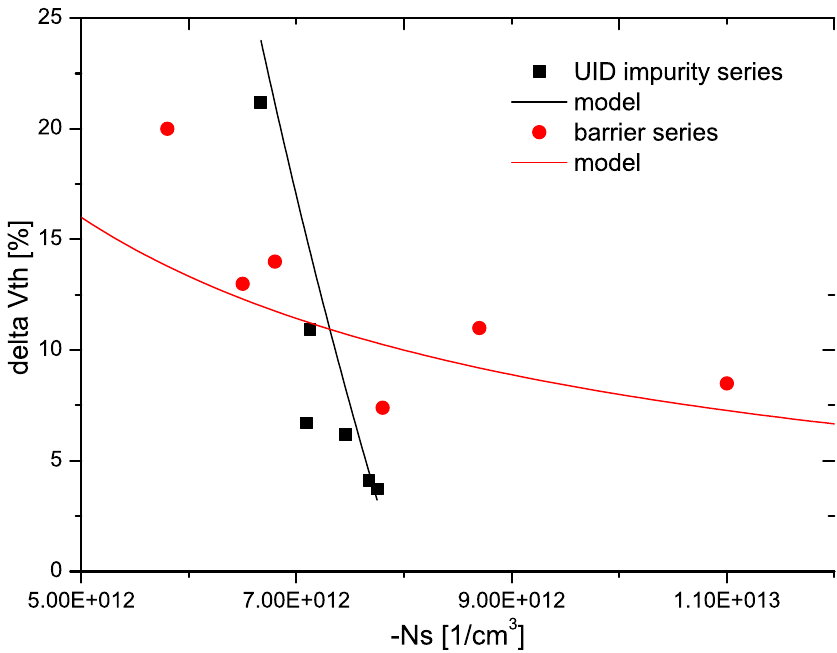}
\caption{V$_{\mathrm{th}}$ shift for a device stressed at V$_{\mathrm{DS}}$=200 V and V$_{\mathrm{GS}}$=-20 V plotted versus the sheet carrier density of the channel. Model: calculated relative V$_{\mathrm{th}}$ shift for varying N$_{\mathrm{s}}$ and N$_{\mathrm{trap}}$ values.}
\label{FIG.7}
\end{figure}

As a first approximation in both cases it is assumed that trap centers do not change independently of the value of N$_{\mathrm{s}}$. The relative values of the V$_{\mathrm{th}}$ shift can be evaluated for both situations and are shown in Fig.\ref{FIG.7}. It can be seen that this approximation is in rough agreement with the experimental data of the UID impurity split and of the barrier split. Furthermore, it needs to be mentioned, that the distance of the traps or possible deep buffer effects cannot be taken into account within this simple model and therefore it is not applicable to understand the results from the UID thickness series.

\begin{figure}[t]
\centering
\includegraphics[scale=0.6]{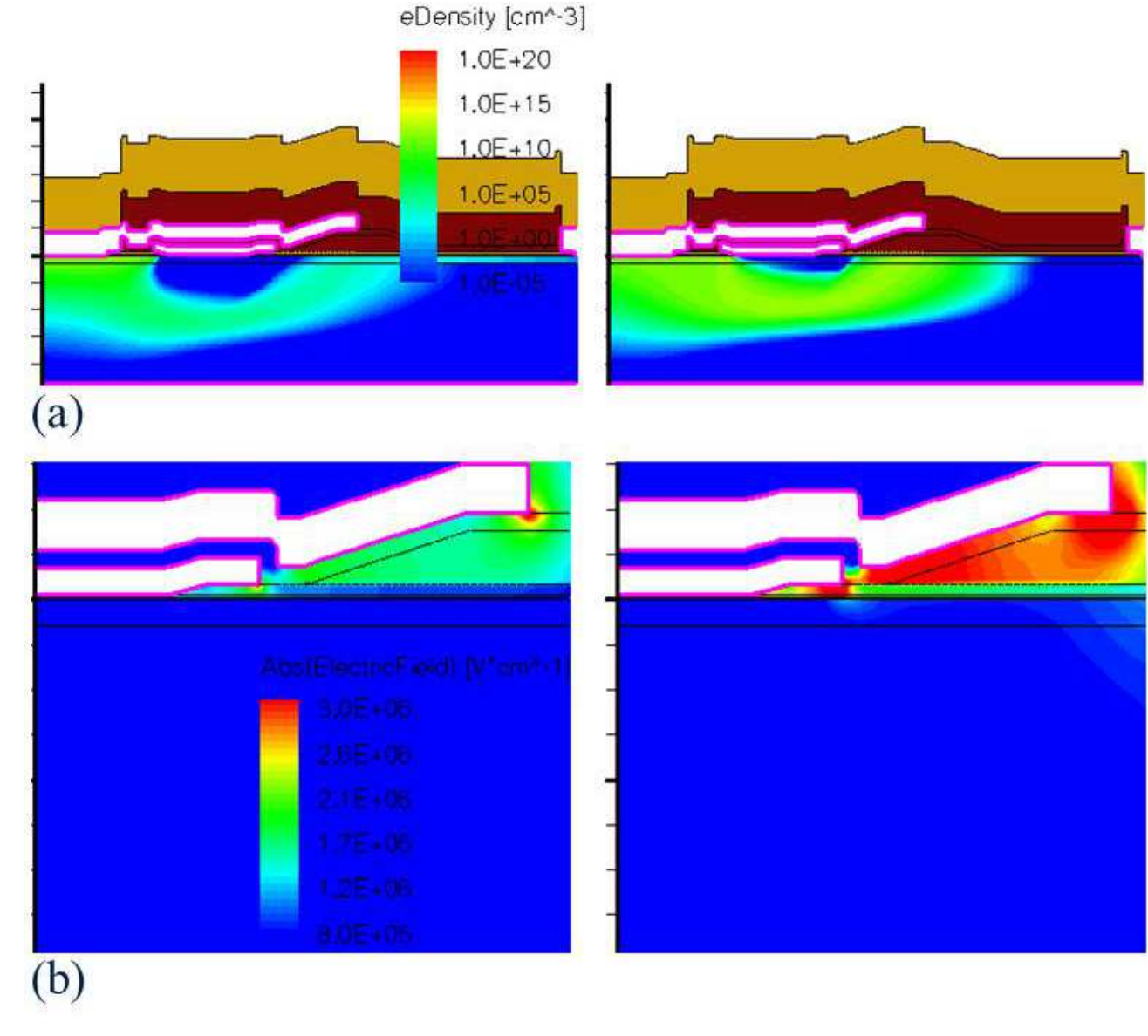}
\caption{(a) 2D color plot section of the electrictron density in the device (left: low N$_{\mathrm{s}}$, right: high N$_{\mathrm{s}}$)  (b) (a) 2D color plot section of the electric field in the device (left: low N$_{\mathrm{s}}$, right: high N$_{\mathrm{s}}$) }
\label{FIG.8}
\end{figure}

\subsubsection{Electric field simulation}
Electric field simulations for different N$_{\mathrm{s}}$ values have been performed with the Synopsys Sentaurus device TCAD tool \cite{Synopsys2012}. Two case studies are considered: (i) Fixed V$_{\mathrm{GS}}$ value in the OFF-state stress (-12 V) independently of the N$_{\mathrm{s}}$ values and (ii) a variable V$_{\mathrm{GS}}$ value in the OFF-state stress in order to compensate for the difference in V$_{\mathrm{th}}$ arising from the different value of N$_{\mathrm{s}}$. The main goal is to assess the impact of the different driving schemes on the vertical electric field in the gate region. As it can be seen in the 2DEG colour plot of the electron density in Fig.\ref{FIG.8}(a), a high value of the N$_{\mathrm{s}}$ shows lower depletion. This could possibly lead to the situation, that due to changed OFF-state stress conditions a lower number of trap centers can be accessed. However, what emerges from simulation results is that the overall electric field is mainly driven by the 2DEG depletion during OFF-state stress condition, which is controlled entirely by the 2DEG density. This means that, independently on the different driving scheme, i.e. fixed or variable gate overdrive in OFF-state conditions, the electric field will increase with increasing N$_{\mathrm{s}}$ (Fig.\ref{FIG.8}(b)). These conclusions have important implication when designing a power MIS-HEMT since higher N$_{\mathrm{s}}$ will provide less dynamic instabilties but, at the same time, will induce a significant increase of the maximum electric field in the different regions of the transistors, with possible important implications on the overall reliability of the GaN device. The results of the simulation show that the 2DEG density, the field plate configuration and the electric field in the MIS-HEMT must be all optimized at the same time in order to ensure sufficiently low dynamic instabilties and a reasonable maximum electric field, especially in the AlGaN barrier overlayer in order to meet the required device lifetime.

\subsection{PL measurement results}
\begin{figure}[]
\centering
\includegraphics[scale=0.95]{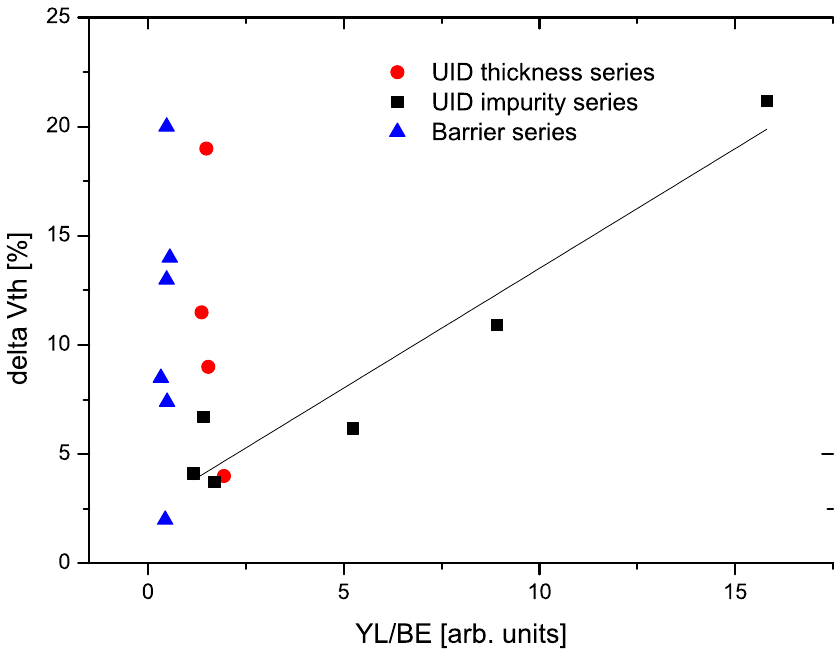}
\caption{V$_{\mathrm{th}}$ shift for a device stressed at V$_{\mathrm{DS}}$=200 V and V$_{\mathrm{GS}}$=-20 V plotted on top of the YL/BE ratio based on room temperature PL spectra.}
\label{FIG.9}
\end{figure}

Steady state PL intensity measured at room temperature (RT) has been performed for all samples sets in this work. The PL spectra from the UID impurity series are dominated by a broad YL band at $\sim$2.2 eV and the band edge (BE) emission at $\sim$3.4 eV. In order to compare  the different sample sets with respect to the YL and by the band edge (BE) emission, peak intensity integration for each of the two peaks has been carried out and shown in Fig.\ref{FIG.9}. Remarkably, a linear correlation between the C concentration and the YL/BE ratio was observed in the UID quality series, indicating that the YL mainly originates from carbon-related defects in the UID-GaN layer. Samples F, H, I and J have been investigated and the results suggest, that the C$_{\mathrm{N}}$O$_{\mathrm{N}}$ complex is the origin of the YL and the main trapping processes available in the hetero structures \cite{Huber2015}. This observed relationship between the YL and the C concentration is in agreement with a recent comprehensive study \cite{Reshchikov2014}. In the other two series' the YL/BE ratio is constant at a very low value. This reflects the fact that the YL/BE ratio in the measurement conditions applied, is a direct result of the C concentration in the UID-GaN layer while the barrier structure and the UID-GaN layer thickness are found to have a minor influence on this ratio. 

\section{Summary and conclusion}

In conclusion, the dynamic instabilities of various identically processed MIS-HEMT structures with different growth conditions for the top epitaxial structure have been investigated by means of SIMS, PL and electric device measurements. Our results suggest that trapping phenomena can be changed in a broad range only by structural and compositional variation during the MOVPE growth process. Moreover, we demonstrate that unintentional impurities like C close to the channel play a major role in V$_{\mathrm{th}}$ shift of the MIS-HEMT devices. It is worth to mention that both the C concentration in the UID-GaN layer and the distance from the channel to the higher doped semi-insulating multilayer structure structure have a direct effect on the V$_{\mathrm{th}}$ shift in MIS-HEMT devices. This implies that the dynamic instabilities are also affected by the amount of C in the semi-insulating multilayer structure layers, either in terms of OFF-state depletion into this area or in terms of absolute thickness of this high C doped region. The results suggest the trapping mechanism itself is dominantely linked to C related defects trapped in the OFF-state conditions. Additionally, we note that experiments to study trapping behavior in AlGaN/GaN MIS-HEMTs should avoid large variations of N$_{\mathrm{s}}$, as this change in N$_{\mathrm{s}}$ itself causes a difference in the V$_{\mathrm{th}}$ shift. The influence of the 2DEG properties to the OFF-state trapping in a MIS-HEMT device is reported for the first time is supported by a basic model and electric field simulation. As long as the layer structure is not changed, the YL/BE ratio is demonstrated to be a powerful tool in predicting trap concentrations in the top part of the MIS-HEMT structures leading to dynamic instabilities, which is also in agreement with earlier works \cite{Fujimoto2008}.


%
\bibliography{library1}

\end{document}